
\input hyperbasics

\input amssym.tex 

\def\unredoffs{}
\tolerance=1000\hfuzz=2pt
\catcode`\@=11 
\ifx\hyperdef\UNd@FiNeD\def\hyperdef#1#2#3#4{#4}\def\hyperref#1#2#3#4{#4}\def\href#1#2{#2}\fi
\magnification=1200\unredoffs\baselineskip=16pt plus 2pt minus 1pt
\def\Date#1{\vfill\leftline{#1}\tenpoint\supereject%
\footline={\hss\tenrm\hyperdef\hypernoname{page}\folio\folio\hss}}%

{\count255=\time\divide\count255 by 60 \xdef\hourmin{\number\count255}
 \multiply\count255 by-60\advance\count255 by\time
 \xdef\hourmin{\hourmin:\ifnum\count255<10 0\fi\the\count255}
}
\def\date{\number\day.\number\month.\number\year\ at \hourmin}


\def\nolabels{\def\wrlabeL##1{}\def\eqlabeL##1{}\def\reflabeL##1{}}
\def\writelabels{\def\wrlabeL##1{\leavevmode\vadjust{\rlap{\smash%
{\line{{\escapechar=` \hfill\rlap{\sevenrm\hskip.03in\string##1}}}}}}}%
\def\eqlabeL##1{{\escapechar-1\rlap{\sevenrm\hskip.05in\string##1}}}%
\def\reflabeL##1{\noexpand\llap{\noexpand\sevenrm\string\string\string##1}}}
\nolabels

\global\newcount\secno \global\secno=0
\global\newcount\meqno \global\meqno=1
\def\s@csym{}

\def\newsec#1\par{\global\advance\secno by1%
{\toks0{#1}\message{(\the\secno. \the\toks0)}}%
\global\subsecno=0\eqnres@t\let\s@csym\secsym\xdef\secn@m{\the\secno}\noindent
{\bf\hyperdef\hypernoname{section}{\the\secno}{\the\secno.} #1}%
\writetoca{{\string\hyperref{}{section}{\the\secno}{\bf \the\secno\quad}} {\bf #1}}\par%
\nobreak\medskip\nobreak\noindent\ignorespaces}
\def\eqnres@t{\xdef\secsym{\the\secno.}\global\meqno=1\bigbreak\bigskip}
\def\sequentialequations{\def\eqnres@t{\bigbreak}}\xdef\secsym{}

\global\newcount\subsecno \global\subsecno=0
\def\subsec#1\par{\global\advance\subsecno by1%
{\toks0{#1}\message{(\s@csym\the\subsecno. \the\toks0)}}%
\global\subsubsecno=0%
\ifnum\lastpenalty>9000\else\bigbreak\fi
\noindent{\it\hyperdef\hypernoname{subsection}{\secn@m.\the\subsecno}%
{\secn@m.\the\subsecno.} #1}\writetoca{\string\hskip1.45cm
{\string\hyperref{}{subsection}{\secn@m.\the\subsecno}{\secn@m.\the\subsecno.}}
{#1}}\par\nobreak\medskip\nobreak\noindent\ignorespaces}

\global\newcount\subsubsecno \global\subsubsecno=0
\def\subsubsec#1\par{\global\advance\subsubsecno by1%
{\toks0{#1}\message{(\secn@m.\the\subsecno.\the\subsubsecno. \the\toks0)}}%
\global\subsubsubsecno=0%
\ifnum\lastpenalty>9000\else\bigbreak\fi
\noindent{\it\hyperdef\hypernoname{subsubsection}{\secn@m.\the\subsecno\the\subsubsecno}%
{\secn@m.\the\subsecno.\the\subsubsecno.} #1}
\par\nobreak\medskip\nobreak\noindent\ignorespaces}

\global\newcount\subsubsubsecno \global\subsubsubsecno=0
\def\subsubsubsec#1\par{\global\advance\subsubsubsecno by1%
{\toks0{#1}\message{(\secn@m.\the\subsecno.\the\subsubsecno.\the\subsubsubsecno \the\toks0)}}%
\ifnum\lastpenalty>9000\else\bigbreak\fi
\noindent{\it\hyperdef\hypernoname{subsubsection}{\secn@m.\the\subsecno\the\subsubsecno\the\subsubsubsecno}%
{\secn@m.\the\subsecno.\the\subsubsecno.\the\subsubsubsecno.} #1}%
\par\nobreak\medskip\nobreak\noindent\ignorespaces}


\def\newnewsec#1#2\par{\global\advance\secno by1%
{\toks0{#2}\message{(\secn@m. \the\toks0)}}%
\global\subsecno=0\global\subsubsecno=0\eqnres@t\let\s@csym\secsym\xdef\secn@m{\the\secno}\noindent
\ifnum\lastpenalty>9000\else\bigbreak\fi
\noindent{\bf\hyperdef\hypernoname{section}{\secn@m}{\secn@m.} #2}%
\writetoca{{\string\hyperref{}{section}{\the\secno}{\bf \the\secno\quad}} {\bf #2}}
\DefWarn#1%
\xdef#1{\noexpand\hyperref{}{section}{\the\secno}%
{\the\secno}}\writedef{#1\leftbracket#1}\wrlabeL{#1=#1}%
\par\nobreak\medskip\nobreak\noindent\ignorespaces}

\def\newsubsec#1#2\par{\global\advance\subsecno by1%
{\toks0{#2}\message{(\secn@m.\the\subsecno. \the\toks0)}}%
\global\subsubsecno=0%
\ifnum\lastpenalty>9000\else\bigbreak\fi
\noindent{\it\hyperdef\hypernoname{subsection}{\secn@m.\the\subsecno}%
{\secn@m.\the\subsecno.} #2}
\DefWarn#1%
\xdef#1{\noexpand\hyperref{}{subsection}{\secn@m.\the\subsecno}%
{\secn@m.\the\subsecno}}\writedef{#1\leftbracket#1}\wrlabeL{#1=#1}%
\writetoca{\string\hskip1.45cm
{\string\hyperref{}{subsection}{\secn@m.\the\subsecno}{\secn@m.\the\subsecno.}}
{#2}}%
\par\nobreak\medskip\nobreak\noindent\ignorespaces}

\def\newsubsecstar#1#2\par{\global\advance\subsecno by1%
{\toks0{#2}\message{(\secn@m.\the\subsecno. \the\toks0)}}%
\global\subsubsecno=0%
\ifnum\lastpenalty>9000\else\bigbreak\fi
\noindent{\it\hyperdef\hypernoname{subsection}{\secn@m.\the\subsecno}%
{\secn@m.\the\subsecno.} #2}
\DefWarn#1%
\xdef#1{\noexpand\hyperref{}{subsection}{\secn@m.\the\subsecno}%
{\secn@m.\the\subsecno}}\writedef{#1\leftbracket#1}\wrlabeL{#1=#1}%
\par\nobreak\medskip\nobreak\noindent\ignorespaces}

\def\newsubsubsec#1#2\par{\global\advance\subsubsecno by1%
{\toks0{#2}\message{(\secn@m.\the\subsecno.\the\subsubsecno. \the\toks0)}}%
\global\subsubsubsecno=0%
\ifnum\lastpenalty>9000\else\bigbreak\fi
\noindent{\it\hyperdef\hypernoname{subsubsection}{\secn@m.\the\subsecno.\the\subsubsecno}%
{\secn@m.\the\subsecno.\the\subsubsecno.} #2}
\DefWarn#1%
\xdef#1{\noexpand\hyperref{}{subsubsection}{\secn@m.\the\subsecno.\the\subsubsecno}%
{\secn@m.\the\subsecno.\the\subsubsecno}}\writedef{#1\leftbracket#1}\wrlabeL{#1=#1}%
\par\nobreak\medskip\nobreak\noindent\ignorespaces}

\def\newsubsubsubsec#1#2\par{\global\advance\subsubsubsecno by1%
{\toks0{#2}\message{(\secn@m.\the\subsecno.\the\subsubsecno.\the\subsubsubsecno \the\toks0)}}%
\ifnum\lastpenalty>9000\else\bigbreak\fi
\noindent{\it\hyperdef\hypernoname{subsubsection}{\secn@m.\the\subsecno\the\subsubsecno\the\subsubsubsecno}%
{\secn@m.\the\subsecno.\the\subsubsecno.\the\subsubsubsecno.} #2}
\DefWarn#1%
\xdef#1{\noexpand\hyperref{}{subsubsubsection}{\secn@m.\the\subsecno.\the\subsubsecno.\the\subsubsubsecno}%
{\secn@m.\the\subsecno.\the\subsubsecno.\the\subsubsubsecno}}\writedef{#1\leftbracket#1}\wrlabeL{#1=#1}%
\par\nobreak\medskip\nobreak\noindent\ignorespaces}

\def\appendix#1#2{\global\meqno=1\global\subsecno=0\global\subsubsecno=0\xdef\secsym{\hbox{#1.}}%
\bigbreak\bigskip\noindent{\bf Appendix \hyperdef\hypernoname{appendix}{#1}%
{#1.} #2}{\toks0{(#1. #2)}\message{\the\toks0}}%
\xdef\s@csym{#1.}\xdef\secn@m{#1}%
\writetoca{{\string\hyperref{}{appendix}{#1}{\bf {#1}\quad}} {\bf #2}}%
\par\nobreak\medskip\nobreak}

%
\def\checkm@de#1#2{\ifmmode{\def\f@rst##1{##1}\hyperdef\hypernoname{equation}%
{#1}{#2}}\else\hyperref{}{equation}{#1}{#2}\fi}
\def\eqnn#1{\DefWarn#1\xdef #1{(\noexpand\relax\noexpand\checkm@de%
{\s@csym\the\meqno}{\secsym\the\meqno})}%
\wrlabeL#1\writedef{#1\leftbracket#1}\global\advance\meqno by1}
\def\f@rst#1{\c@t#1a\em@ark}\def\c@t#1#2\em@ark{#1}
\def\eqna#1{\DefWarn#1\wrlabeL{#1$\{\}$}%
\xdef #1##1{(\noexpand\relax\noexpand\checkm@de%
{\s@csym\the\meqno\noexpand\f@rst{##1}1}{\hbox{$\secsym\the\meqno##1$}})}
\writedef{#1\numbersign1\leftbracket#1{\numbersign1}}\global\advance\meqno by1}
\def\eqn#1#2{\DefWarn#1%
\xdef #1{(\noexpand\hyperref{}{equation}{\s@csym\the\meqno}%
{\secsym\the\meqno})}$$#2\eqno(\hyperdef\hypernoname{equation}%
{\s@csym\the\meqno}{\secsym\the\meqno})\eqlabeL#1$$%
\writedef{#1\leftbracket#1}\global\advance\meqno by1}
\def\xeqn{\expandafter\xe@n}\def\xe@n(#1){#1}
\def\xeqna#1{\expandafter\xe@n#1}
\def\eqns#1{(\e@ns #1{\hbox{}})}
\def\e@ns#1{\ifx\UNd@FiNeD#1\message{eqnlabel \string#1 is undefined.}%
\xdef#1{(?.?)}\fi{\let\hyperref=\relax\xdef\next{#1}}%
\ifx\next\em@rk\def\next{}\else%
\ifx\next#1\xeqn#1\else\def\n@xt{#1}\ifx\n@xt\next#1\else\xeqna#1\fi
\fi\let\next=\e@ns\fi\next}
\def\DefWarn#1{}
%
\newskip\footskip\footskip14pt plus 1pt minus 1pt 
\def\footnotefont{\ninepoint}\def\f@t#1{\footnotefont #1\@foot}
\def\f@@t{\baselineskip\footskip\bgroup\footnotefont\aftergroup\@foot\let\next}
\setbox\strutbox=\hbox{\vrule height9.5pt depth4.5pt width0pt}
\global\newcount\ftno \global\ftno=0
\def\foot{\global\advance\ftno by1\def\foot@rg{\hyperref{}{footnote}%
{\the\ftno}{\the\ftno}\xdef\foot@rg{\noexpand\hyperdef\noexpand\hypernoname%
{footnote}{\the\ftno}{\the\ftno}}}\footnote{$^{\foot@rg}$}}
%
%
%
\global\newcount\refno \global\refno=1
\newwrite\rfile
\def\ref{[\hyperref{}{reference}{\the\refno}{\the\refno}]\nref}
\def\nref#1{\DefWarn#1%
\xdef#1{[\noexpand\hyperref{}{reference}{\the\refno}{\the\refno}]}%
\writedef{#1\leftbracket#1}%
\ifnum\refno=1\immediate\openout\rfile=\jobname.refs\fi
\chardef\wfile=\rfile\immediate\write\rfile{\noexpand\item{[\noexpand\hyperdef%
\noexpand\hypernoname{reference}{\the\refno}{\the\refno}]\ }%
\reflabeL{#1\hskip.31in}\pctsign}\global\advance\refno by1\findarg}
\def\findarg#1#{\begingroup\obeylines\newlinechar=`\^^M\pass@rg}
{\obeylines\gdef\pass@rg#1{\writ@line\relax #1^^M\hbox{}^^M}%
\gdef\writ@line#1^^M{\expandafter\toks0\expandafter{\striprel@x #1}%
\edef\next{\the\toks0}\ifx\next\em@rk\let\next=\endgroup\else\ifx\next\empty%
\else\immediate\write\wfile{\the\toks0}\fi\let\next=\writ@line\fi\next\relax}}
\def\striprel@x#1{} \def\em@rk{\hbox{}}
\def\lref{\begingroup\obeylines\lr@f}
\def\lr@f#1#2{\DefWarn#1\gdef#1{\let#1=\UNd@FiNeD\ref#1{#2}}\endgroup\unskip}
\def\semi{;\hfil\break}
\def\addref#1{\immediate\write\rfile{\noexpand\item{}#1}} 
\def\listrefs{\vfill\supereject\immediate\closeout\rfile\writestoppt
\baselineskip=\footskip\centerline{{\bf References}}\bigskip{\parindent=20pt%
\frenchspacing\escapechar=` \input \jobname.refs\vfill\eject}\nonfrenchspacing}
\def\startrefs#1{\immediate\openout\rfile=\jobname.refs\refno=#1}
\def\xref{\expandafter\xr@f}\def\xr@f[#1]{#1}
\def\refs#1{\count255=1[\r@fs #1{\hbox{}}]}
\def\r@fs#1{\ifx\UNd@FiNeD#1\message{reflabel \string#1 is undefined.}%
\nref#1{need to supply reference \string#1.}\fi%
\vphantom{\hphantom{#1}}{\let\hyperref=\relax\xdef\next{#1}}%
\ifx\next\em@rk\def\next{}%
\else\ifx\next#1\ifodd\count255\relax\xref#1\count255=0\fi%
\else#1\count255=1\fi\let\next=\r@fs\fi\next}
%

%
\newwrite\ffile\global\newcount\figno \global\figno=1
\def\fig{fig.~\hyperref{}{figure}{\the\figno}{\the\figno}\nfig}
\def\nfig#1{\DefWarn#1%
\xdef#1{fig.~\noexpand\hyperref{}{figure}{\the\figno}{\the\figno}}%
\writedef{#1\leftbracket fig.\noexpand~\xfig#1}%
\ifnum\figno=1\immediate\openout\ffile=\jobname.figs\fi\chardef\wfile=\ffile%
{\let\hyperref=\relax
\immediate\write\ffile{\noexpand\medskip\noexpand\item{Fig.\ %
\noexpand\hyperdef\noexpand\hypernoname{figure}{\the\figno}{\the\figno}. }
\reflabeL{#1\hskip.55in}\pctsign}}\global\advance\figno by1\findarg}
\def\xfig{\expandafter\xf@g}\def\xf@g fig.\penalty\@M\ {}
\def\figs#1{figs.~\f@gs #1{\hbox{}}}
\def\f@gs#1{{\let\hyperref=\relax\xdef\next{#1}}\ifx\next\em@rk\def\next{}\else
\ifx\next#1\xfig #1\else#1\fi\let\next=\f@gs\fi\next}
%
\def\figin{\epsfcheck\figin}\def\figins{\epsfcheck\figins}
\def\epsfcheck{\ifx\epsfbox\UnDeFiNeD
\message{(NO epsf.tex, FIGURES WILL BE IGNORED)}
\gdef\figin##1{\vskip2in}\gdef\figins##1{\hskip.5in}
\else\message{(FIGURES WILL BE INCLUDED)}%
\gdef\figin##1{##1}\gdef\figins##1{##1}\fi}
\def\figinsert{\goodbreak\topinsert}
\def\ifig#1#2#3{\DefWarn#1\xdef#1{fig.~\the\figno}
\writedef{#1\leftbracket fig.\noexpand~\the\figno}%
\figinsert\figin{\centerline{#3}}
\smallskip
\leftskip=0pt \rightskip=0pt
\baselineskip12pt\noindent
{{\bf Fig.~\the\figno}\ \ninepoint #2}
\medskip
\global\advance\figno by1\par\endinsert}
\newwrite\lfile
{\escapechar-1\xdef\pctsign{\string\%}\xdef\leftbracket{\string\{}
\xdef\rightbracket{\string\}}\xdef\numbersign{\string\#}}
\def\writedefs{\immediate\openout\lfile=label.defs \def\writedef##1{%
{\let\hyperref=\relax\let\hyperdef=\relax\let\hypernoname=\relax
 \immediate\write\lfile{\string\checkdef\string##1\rightbracket}}}}%
\def\writestop{\def\writestoppt{\immediate\write\lfile{\string\pageno
 \the\pageno\string\startrefs\leftbracket\the\refno\rightbracket
 \string\def\string\secsym\leftbracket\secsym\rightbracket
 \string\secno\the\secno\string\meqno\the\meqno}\immediate\closeout\lfile}}
\def\writestoppt{}\def\writedef#1{}

\def\seclab#1\par{\DefWarn#1%
\xdef #1{\noexpand\hyperref{}{section}{\the\secno}{\the\secno}}%
\writedef{#1\leftbracket#1}\wrlabeL{#1=#1}\par%
\nobreak\medskip\nobreak\noindent\ignorespaces}
\def\subseclab#1\par{\DefWarn#1%
\xdef #1{\noexpand\hyperref{}{subsection}{\the\secno.\the\subsecno}%
{\the\secno.\the\subsecno}}\writedef{#1\leftbracket#1}\wrlabeL{#1=#1}\par%
\nobreak\medskip\nobreak\noindent\ignorespaces}
\def\subsubseclab#1\par{\DefWarn#1%
\xdef#1{\noexpand\hyperref{}{subsubsection}{\the\secno.\the\subsecno.\the\subsubsecno}%
{\the\secno.\the\subsecno.\the\subsubsecno}}\writedef{#1\leftbracket#1}\wrlabeL{#1=#1}\par%
\nobreak\medskip\nobreak\noindent\ignorespaces}
\def\applab#1\par{\DefWarn#1%
\xdef#1{\noexpand\hyperref{}{appendix}{\secn@m}{\secn@m}}%
\writedef{#1\leftbracket#1}\wrlabeL{#1=#1}%
\par\nobreak\medskip\nobreak\noindent\ignorespaces}
\def\appsublab#1{\DefWarn#1%
\xdef #1{\noexpand\hyperref{}{appendix}{\secn@m.\the\subsecno}{\secn@m.\the\subsecno}}%
\writedef{#1\leftbracket#1}\wrlabeL{#1=#1}}
\newwrite\tfile \def\writetoca#1{}
\def\leaderfill{\leaders\hbox to 1em{\hss.\hss}\hfill}
\def\writetoc{\immediate\openout\tfile=\jobname.toc
   \def\writetoca##1{{\edef\next{\write\tfile{\noindent ##1
   \string\leaderfill{
   \string\hyperref{}{page}{\noexpand\number\pageno}%
   {\noexpand\number\pageno}} \par}}\next}}
}
\newread\ch@ckfile
\def\listtoc{\immediate\closeout\tfile\immediate\openin\ch@ckfile=\jobname.toc
\ifeof\ch@ckfile\message{no file \jobname.toc, no table of contents this pass}%
\else\closein\ch@ckfile\centerline{\bf Contents}\nobreak\medskip%
{\baselineskip=15.5pt\footnotefont\parskip=0pt\catcode`\@=11\input\jobname.toc
\catcode`\@=12\bigbreak\bigskip}\fi}
\catcode`\@=12 
\def\tenpoint{\def\rm{\fam0\tenrm}
\textfont0=\tenrm \scriptfont0=\sevenrm \scriptscriptfont0=\fiverm
\textfont1=\teni  \scriptfont1=\seveni  \scriptscriptfont1=\fivei
\textfont2=\tensy \scriptfont2=\sevensy \scriptscriptfont2=\fivesy
\textfont\itfam=\tenit \def\it{\fam\itfam\tenit}\def\footnotefont{\ninepoint}%
\textfont\bffam=\tenbf \def\bf{\fam\bffam\tenbf}\def\sl{\fam\slfam\tensl}\rm}
\font\ninerm=cmr9 \font\sixrm=cmr6 \font\ninei=cmmi9 \font\sixi=cmmi6
\font\ninesy=cmsy9 \font\sixsy=cmsy6 \font\ninebf=cmbx9
\font\nineit=cmti9 \font\ninesl=cmsl9 \skewchar\ninei='177
\skewchar\sixi='177 \skewchar\ninesy='60 \skewchar\sixsy='60
\def\ninepoint{\def\rm{\fam0\ninerm}
\textfont0=\ninerm \scriptfont0=\sixrm \scriptscriptfont0=\fiverm
\textfont1=\ninei \scriptfont1=\sixi \scriptscriptfont1=\fivei
\textfont2=\ninesy \scriptfont2=\sixsy \scriptscriptfont2=\fivesy
\textfont\itfam=\ninei \def\it{\fam\itfam\nineit}\def\sl{\fam\slfam\ninesl}%
\textfont\bffam=\ninebf \def\bf{\fam\bffam\ninebf}\rm}
%
\hyphenation{anom-aly anom-alies coun-ter-term coun-ter-terms}

\def\tikzcaption#1#2{\DefWarn#1\xdef#1{Fig.~\the\figno}
\writedef{#1\leftbracket Fig.\noexpand~\the\figno}%
{
\smallskip
\leftskip=20pt \rightskip=20pt \baselineskip12pt\noindent
{{\bf Fig.~\the\figno}\ \ninepoint #2}
\bigskip
\global\advance\figno by1 \par}}

\def\ntoalpha#1{%
\ifcase#1%
@%
\or A\or B\or C\or D\or E\or F\or G\or H\or I\or J\or K\or L\or M%
\fi
}

\global\newcount\appno \global\appno=1
\def\applab#1{\xdef #1{\ntoalpha{\appno}}\writedef{#1\leftbracket#1}\wrlabeL{#1=#1}
\global\advance\appno by1}

\def\preprint#1 #2\par{\rightline{\vbox{\baselineskip12pt\hbox{#1}\hbox{#2}}}\vskip2cm}
%
\def\title#1\par{\centerline{\bf #1}\nopagenumbers\pageno=0}
\def\author#1\par{\bigskip\bigskip\centerline{#1}}

\newcount\addressno

\def\email#1#2{
\footnote{\null}{\kern-\parindent \llap{$^#1$\hskip1pt}email: #2}}

\def\startcenter{%
  \par
  \begingroup
  \leftskip=0pt plus 1fil
  \rightskip=\leftskip
  \parindent=0pt
  \parfillskip=0pt
}
\def\stopcenter{\endgroup}

\def\address{\bigskip%
  \ifnum\the\addressno=0\else\stopcenter\endgroup\fi
  \advance\addressno by 1%
  \begingroup
  \startcenter
  \it
  \obeylines
  \addressAux
}
\def\addressAux#1{#1}

\def\abstract{\stopcenter\endgroup\bigskip\bigskip\noindent}

\def\Dsl{\,\raise.15ex\hbox{/}\mkern-13.5mu D} 
\def\dsl{\raise.15ex\hbox{/}\kern-.57em\partial}
 
\def\boxeqn#1{\vcenter{\vbox{\hrule\hbox{\vrule\kern3pt\vbox{\kern3pt
	\hbox{${\displaystyle #1}$}\kern3pt}\kern3pt\vrule}\hrule}}}


\def\ap{{\alpha^{\prime}}}

\def\a{\alpha}
\def\b{{\beta}}
\def\g{{\gamma}}

\def\l{\lambda}

\def\t{{\theta}}

\def\half{{1\over 2}}
\def\p{{\partial}}

\def\({\left(}
\def\){\right)}

\def\cF{{\cal F}}

\def\cW{{\cal W}}



\def\len#1{{%
\def\Dlen{\left|\mkern-1mu #1\mkern -0.5mu\right|}%
\def\Sslen{\left|\mkern-1.3mu #1\mkern -1.3mu\right|}%
\def\SSlen{\left|\mkern-2.8mu #1\mkern-1.3mu\right|}%
\mathchoice{\Dlen}{\Dlen}{\Sslen}{\SSlen}}}

\def\sfrac#1/#2{\kern.1em\raise.5ex\hbox{\the\scriptfont0 #1}%
\kern-.1em/\kern-.15em\lower.25ex\hbox{\the\scriptfont0 #2}}

\font\tenshuffle=shuffle10 \font\sevenshuffle=shuffle7 \font\fiveshuffle=shuffle7 at 5pt
\def\shuffle{{%
\def\Dshuffle{\mathbin{\hbox{\tenshuffle\char'001}}}%
\def\Sshuffle{\mathbin{\hbox{\sevenshuffle\char'001}}}%
\def\SSshuffle{\mathbin{\hbox{\fiveshuffle\char'001}}}%
\mathchoice{\Dshuffle}{\Dshuffle}{\Sshuffle}{\SSshuffle}}}


\def\qed{\hbox{\hskip 3pt
\vbox{\hrule\hbox to 7pt{\vrule height 7pt\hfill\vrule}
\hrule}}\hskip3pt}

\overfullrule=0pt\relax

\frenchspacing

\def\checkdef#1#2{
\ifx\UndeFined#1%
	\def#1{#2}
\else
	\immediate\write16{*** BUG ***: the label \string#1 is already defined ***}
\fi
}
\newread\instream

\openin\instream= label.defs
\ifeof\instream\message{No labels in advance yet. Wait till next pass.}
\else\closein\instream \input label.defs
\fi
\writedefs

\def\arXiv:#1].{\hepthStrip#1 \nil}
\def\hepthStrip#1 #2\nil{\href{http://arxiv.org/abs/#1}{arXiv:#1 #2\unskip}].}

\font\frakfont=eufm10 at 10pt
\def\ce{\mathord{\hbox{\frakfont e}}}
\def\cf{\mathord{\hbox{\frakfont f}}}

\input amssym

\def\textbf#1{{\bf #1}}
\def\paragraph#1{\medskip\noindent{\it #1.}}
\def\cX{{\cal X}}


\title Towards massive field-theory amplitudes

\title from the cohomology of pure spinor superspace

\author
Carlos R. Mafra\email{{}}{c.r.mafra@soton.ac.uk}

\address
Mathematical Sciences and STAG Research Centre, University of Southampton,
Highfield, Southampton, SO17 1BJ, UK

\abstract
By analogy with the formula for the massless string disk amplitudes, we define
massive field-theory tree amplitudes and conjecture that the BRST cohomology structure of
pure spinor superspace fixes their form. We give evidence by deriving
the pure spinor superspace expression of the massive field-theory
$n$-point tree amplitude with one first-level massive and $n{-}1$ massless
states in two ways:  1) from BRST cohomology arguments in pure spinor superspace and 2) from the $\ap^2$
correction to the massless string amplitudes by inverting the unitarity constraint in superspace.

\Date{August 2024}

\lref\aomoto{
K. Aomoto, J. Math. Soc. Japan 39 (1987) 191-208.
}

\lref\psswebsite{
	http://www.southampton.ac.uk/\~{}crm1n16/pss.html
}

\lref\Ree{
	R.~Ree, ``Lie elements and an algebra associated with shuffles'',
	Ann. Math. {\bf 62}, No. 2 (1958), 210--220.
}
\lref\KKref{
	R.~Kleiss and H.~Kuijf,
	``Multi - Gluon Cross-sections and Five Jet Production at Hadron Colliders,''
	Nucl.\ Phys.\ B {\bf 312}, 616 (1989)..
\semi
	V.~Del Duca, L.J.~Dixon and F.~Maltoni,
	``New color decompositions for gauge amplitudes at tree and loop level,''
	Nucl.\ Phys.\ B {\bf 571}, 51 (2000).
	[hep-ph/9910563].
}

\lref\massivevone{
	S.P.~Kashyap, C.R.~Mafra, M.~Verma and L.A.~Ypanaque,
	``A relation between massive and massless string tree amplitudes,''
	[arXiv:2311.12100 [hep-th]].
}
\lref\nptMethod{
	C.~R.~Mafra, O.~Schlotterer, S.~Stieberger and D.~Tsimpis,
	``A recursive method for SYM n-point tree amplitudes,''
	Phys.\ Rev.\ D {\bf 83}, 126012 (2011).
	[arXiv:1012.3981 [hep-th]].
}
\lref\towards{
	C.R.~Mafra,
	``Towards Field Theory Amplitudes From the Cohomology of Pure Spinor Superspace,''
	JHEP {\bf 1011}, 096 (2010).
	[arXiv:1007.3639 [hep-th]].
}
\lref\Soares{
	B.~R.~Soares,
	``Constructing massive superstring vertex operators from massless vertex operators using the pure spinor formalism,''
	Phys. Lett. B \textbf{852}, 138611 (2024)
	[arXiv:2401.03208 [hep-th]].
}

\lref\website{
	http://www.southampton.ac.uk/\~{}crm1n16/pss.html
}
\lref\EOMbbs{
	C.R.~Mafra and O.~Schlotterer,
  	``Multiparticle SYM equations of motion and pure spinor BRST blocks,''
	JHEP {\bf 1407}, 153 (2014).
	[arXiv:1404.4986 [hep-th]].
}
\lref\BGap{
	C.R.~Mafra and O.~Schlotterer,
  	``Non-abelian $Z$-theory: Berends-Giele recursion for the $\alpha'$-expansion of disk integrals,''
	[arXiv:1609.07078].
	{\tt http://repo.or.cz/BGap.git}
}

\lref\partIcohomology{
	C.R.~Mafra and O.~Schlotterer,
	``Cohomology foundations of one-loop amplitudes in pure spinor superspace,''
	[arXiv:1408.3605 [hep-th]].
}

\lref\cdescent{
	C.R.~Mafra,
	``KK-like relations of $\alpha$' corrections to disk amplitudes,''
	JHEP \textbf{03}, 012 (2022)
	[arXiv:2108.01081 [hep-th]].
}

\lref\massSweden{
       	M.~Guillen, H.~Johansson, R.~L.~Jusinskas and O.~Schlotterer,
	``Scattering Massive String Resonances through Field-Theory Methods,''
	Phys. Rev. Lett. \textbf{127}, no.5, 051601 (2021)
	[arXiv:2104.03314 [hep-th]].
}
\lref\oneloopbb{
	C.R.~Mafra and O.~Schlotterer,
	``The Structure of n-Point One-Loop Open Superstring Amplitudes,''
	JHEP \textbf{08}, 099 (2014)
	[arXiv:1203.6215 [hep-th]].
}

\lref\MSSI{
	C.R.~Mafra, O.~Schlotterer and S.~Stieberger,
	``Complete N-Point Superstring Disk Amplitude I. Pure Spinor Computation,''
	Nucl.\ Phys.\ B {\bf 873}, 419 (2013).
	[arXiv:1106.2645 [hep-th]].
}
\lref\MSSII{
	C.~R.~Mafra, O.~Schlotterer and S.~Stieberger,
	``Complete N-Point Superstring Disk Amplitude II. Amplitude
	and Hypergeometric Function Structure,''
	Nucl.\ Phys.\ B {\bf 873}, 461 (2013).
	[arXiv:1106.2646 [hep-th]].
}
\lref\fourtree{
	C.R.~Mafra,
	``Pure Spinor Superspace Identities for Massless Four-point Kinematic Factors,''
	JHEP \textbf{04}, 093 (2008)
	[arXiv:0801.0580 [hep-th]].
}
\lref\PSthreemass{
	S.~Chakrabarti, S.~P.~Kashyap and M.~Verma,
	``Amplitudes Involving Massive States Using Pure Spinor Formalism,''
	JHEP \textbf{12}, 071 (2018)
	[arXiv:1808.08735 [hep-th]].
}
\lref\drinfeld{
	J.~Broedel, O.~Schlotterer, S.~Stieberger and T.~Terasoma,
  	``All order $\alpha^{\prime}$-expansion of superstring trees from the Drinfeld associator,''
	Phys.\ Rev.\ D {\bf 89}, no. 6, 066014 (2014).
	[arXiv:1304.7304 [hep-th]].
}

\lref\masstheta{
	S.~Chakrabarti, S.~P.~Kashyap and M.~Verma,
	``Theta Expansion of First Massive Vertex Operator in Pure Spinor,''
	JHEP \textbf{01}, 019 (2018)
	[arXiv:1706.01196 [hep-th]].
}

\lref\PSS{
	C.R.~Mafra,
	``PSS: A FORM Program to Evaluate Pure Spinor Superspace Expressions,''
	[arXiv:1007.4999 [hep-th]].
}

\lref\ICTP{
	N.~Berkovits,
  	``ICTP lectures on covariant quantization of the superstring,''
	[hep-th/0209059].
}

\lref\BCpaper{
	N.~Berkovits and O.~Chandia,
	``Massive superstring vertex operator in D = 10 superspace,''
	JHEP \textbf{08}, 040 (2002)
	[arXiv:hep-th/0204121 [hep-th]].
}

\lref\psf{
 	N.~Berkovits,
	``Super-Poincare covariant quantization of the superstring,''
	JHEP {\bf 0004}, 018 (2000)
	[arXiv:hep-th/0001035].
}

\lref\wittentwistor{
	E.Witten,
        ``Twistor-Like Transform In Ten-Dimensions''
        Nucl.Phys. B {\bf 266}, 245~(1986)
}
\lref\higherSYM{
	C.R.~Mafra and O.~Schlotterer,
	``A solution to the non-linear equations of D=10 super Yang-Mills theory,''
	Phys.\ Rev.\ D {\bf 92}, no. 6, 066001 (2015).
	[arXiv:1501.05562 [hep-th]].
}
\lref\treereview{
	C.~R.~Mafra and O.~Schlotterer,
	``Tree-level amplitudes from the pure spinor superstring,''
	Phys. Rept. \textbf{1020}, 1-162 (2023)
	[arXiv:2210.14241 [hep-th]].
}
\lref\BGBCJ{
	C.R.~Mafra and O.~Schlotterer,
  	``Berends-Giele recursions and the BCJ duality in superspace and components,''
	JHEP {\bf 1603}, 097 (2016).
	[arXiv:1510.08846 [hep-th]].
}
\lref\Gauge{
	S.~Lee, C.~R.~Mafra and O.~Schlotterer,
	``Non-linear gauge transformations in $D=10$ SYM theory and the BCJ duality,''
	JHEP \textbf{03}, 090 (2016)
	[arXiv:1510.08843 [hep-th]].
}
\lref\FORM{
	J.A.M.~Vermaseren,
	``New features of FORM,''
	arXiv:math-ph/0010025.
\semi
	M.~Tentyukov and J.A.M.~Vermaseren,
	``The multithreaded version of FORM,''
	arXiv:hep-ph/0702279.
}
\lref\MPS{
	N.~Berkovits,
  	``Multiloop amplitudes and vanishing theorems using the pure spinor formalism for the superstring,''
	JHEP {\bf 0409}, 047 (2004).
	[hep-th/0406055].
}

\lref\PSspace{
	N.~Berkovits,
	``Explaining Pure Spinor Superspace,''
	[arXiv:hep-th/0612021 [hep-th]].
}
\lref\UV{
	S.P.~Kashyap, C.R.~Mafra, M.~Verma and L.~Ypanaqu\'e,
	``Massless representation of massive superfields and tree amplitudes with the pure spinor formalism,''
	[arXiv:2407.02436 [hep-th]].
}

\lref\oneloopI{
	C.R.~Mafra and O.~Schlotterer,
	``Towards the n-point one-loop superstring amplitude. Part I. Pure spinors and superfield kinematics,''
	JHEP \textbf{08}, 090 (2019)
	[arXiv:1812.10969 [hep-th]].
}

\listtoc
\writetoc

\newsec{Statement of the problem}

In a recent paper \massSweden, the open superstring disk amplitude
of $n{-}1$ massless states
and one massive state $\underline{n}$ was decomposed in a similar fashion as
the massless amplitudes of \MSSI. Namely, the full string amplitude is written as products of
$(n{-}3)!$ worldsheet integrals $F^P_Q$ independent of polarizations
and partial subamplitudes $A(1,P,n{-}1|\underline{n})$:
\eqn\stringmassS{
{\cal A}(1,Q,n{-}1,\underline{n}) = \sum_{P \in S_{n-3}} F^P_Q
A(1,P,n{-}1|\underline{n})\,.
}
The words $P$ and $Q$ encode the labels of the strings being scattered
while the integrals $F^P_Q$ have the same functional form as
in the massless string scattering amplitude \refs{\MSSI,\MSSII,\drinfeld,\BGap}.

Since the amplitudes $A(1,P|{\underline n})$ play an analogous role as the
massless field-theory
amplitudes $A^{\rm YM}$ in the massless string disk amplitude counterpart of \stringmassS\
given in \MSSI, they will
be called {\it massive field-theory amplitudes}. The same terminology will be
used to expected generalizations of \stringmassS\ with higher number of massive legs and/or
higher mass levels \aomoto\ and should not be understood as taking the $\ap\to0$ limit.

The focus of this paper will be to derive the pure
spinor superspace expression for $A(1,P|{\underline n})$
whose component expansion following
\refs{\psf,\PSspace} reproduces the supersymmetric components found in \massSweden. The pure
spinor superspace expression achieving this,
\eqn\genAuIntro{
A(1,2, \ldots,n{-}1|{\underline n}) = {i\over2\ap}\langle C_{1|2 \ldots n{-}1}^m (\l H^m_n)\rangle\,,
}
will be derived in section~\Camps\ in two different ways:
\smallskip
\item{1.} Finding a BRST closed expression with the correct kinematic pole structure
\item{2.} Inverting the factorization of the massless amplitudes
on their first massive pole

\noindent The first derivation uses the same BRST cohomology ideas \towards\ that were successfully used to
obtain the pure spinor superspace expression of the massless SYM tree-level amplitudes
\nptMethod. The second derivation exploits the relation found in \massivevone\ between the massive
field-theory amplitudes $A(1,P|{\underline n})$ and the $\ap^2$ correction of the disk massless amplitudes.
Both derivations rely on BRST cohomology manipulations in pure spinor superspace.

Similarly to the massless case in \towards, we conjecture that the BRST
cohomology structure of pure spinor superspace fixes the field-theory
massive amplitudes for higher mass levels and higher number of massive legs.
The derivation of \genAuIntro\ here is the first step in this quest.

\newsubsec\revsec Preliminaries

For a review of the pure spinor formalism, we refer the reader to \refs{\ICTP,\treereview}.

\paragraph{Equations of motion} The massless superfields $\big[A_\a, A_m, W^\a, F^{mn}\big]$ \wittentwistor\
and the massive superfields \BCpaper
\eqn\lfields{
\l^\a B_{\a\b}=(\l B)_\b,\quad
\l^\a H^m_\a = (\l H^m)\,,\quad
C^\b{}_\a \l^\a = (C\l)^\b\,,\quad
\l^\a F_{\a mn} = (\l F)_{mn}\,,
}
satisfy the following equations of motion \refs{\wittentwistor,\BCpaper,\UV} (with $A_{[m}B_{n]}=A_m
B_n-A_n B_m$)
\eqn\SYMBRST{
\eqalign{
Q A_\beta +  D_\beta V &= (\g^m\l)_{\beta}A_m\,,\cr
QA_m &= \l\g^m W + \p_m V\,,\cr
Q(\l B)_\a &= (\l\g^m)_\a (\l H)_m\,,\cr
Q(\l H^m) &= (\l\g^m C\l)\,,\cr
}\qquad
\eqalign{
Q W^\a &= {1\over4}(\l\g^{mn})^{\a}F_{mn},\cr
Q F_{mn} &=\p_m(\l\g_n W)- \p_n(\l\g_m W)\,,\cr
Q(C\l)^\a &= {1\over4}(\l\g^{mn})^\a (\l F)_{mn}\,,\cr
Q(\l F)_{mn} &= {1\over2}\p_{[m}(\l\g_{n]} C\l)
-{1\over16}\p^p(\l\g_{[m} C\g_{n]p}\l)\,,
}}
where $Q=\l^\a D_\a$ is the pure spinor BRST operator acting on $10$D superfields, and $D_\a
={\p\over\p\t^a}+\half(\g^m\t)_\a \p_m$ is
the supersymmetric derivative satisfying $\{D_\a,D_\b\}=\g^m_{\a\b}\p_m$. As factors of $\ap$
are being kept in all formulas, we find it convenient to list the length dimension of various
quantities:
\eqnn\lengthdim
$$\displaylines{
[\ap] = 2,\quad [\l^\a]=[\t^\a] = \half,\quad [\p_m] = -1,\quad
[Q] = 0,\quad \hfil\lengthdim\hfilneg\cr
[A_\a] = \half\,,\quad
[A^m] = 0\,,\quad
[W^\a] = -\half\,,\quad
[F^{mn}] = -1\,,\cr
[B_{\a\b}] = 1,\quad
[H_{m\a}] = \half\,,\quad
[C^\b{}_\a] = 0\,,\quad
[F_{\a mn}] = -{1\over2}\,.
}$$

\paragraph{Berends-Giele form of the massive field-theory amplitudes}
Using the perturbiner approach, the supersymmetric partial amplitudes $A(P|\underline{n})$
with $|P|=n{-}1$ massless states represented by gluons $e^m$ and gluinos $\chi^\a$
and one first-level massive state represented by the bosons $g^{mn}, b^{mnp}$
and fermions $\psi^m_{\a}$ (see appendix~\thetaexpsec)
were found to be \massSweden,
\eqn\APun{
A(P|{\underline n}) = \phi_P^{mn} g_{{\underline n}\,mn} + \phi_P^{mnp} b_{{\underline n}\,mnp}
+ \phi_P^{m\a}\psi^m_{{\underline n}\a}\,,
}
where the massless SYM states $(e^m, \chi^\a)$ are encoded in the deconcatenations
\eqnn\allmul
$$\eqalignno{
\phi_P^{mn} &= \sum_{P=XY}\!\!\ap\bigl[ \cf_X^{ma}\cf_Y^{na} + (\cX_X \g^m \cX_Y)k_Y^n\bigr]
 -\!\!\!  \sum_{P=XYZ}\!\!\! 2\ap(\cX_X \g^m \cX_Z)\ce_Y^n + {\rm cyc}(P)\,,\qquad{} &\allmul\cr
\phi_P^{mnp} &= \sum_{P=XY} \bigl[\ce_X^m \ce_Y^n k_Y^p -{1\over12}(\cX_X\g^{mnp}\cX_Y)\bigr]
+ \sum_{P=XYZ}{2\over3}\ce_X^m \ce_Y^n \ce_Z^p + {\rm cyc}(P)\,,\cr
\phi_P^{m\a} &= {2\over9}\ap k_P^n \sum_{P=XY}\cf_X^{mp}(\g_n\g_p\cX_Y)^\a + {\rm cyc}(P)\,,
}$$
whose coefficients are adapted to the conventions of this paper and differ\foot{I thank Oliver
Schlotterer for pointing out the relation $\psi^{m\a} = k^n\g_n^{\a\b}\psi_\b^m$ between the
Weyl fermions $\psi^{m\a}$ of \massSweden\ and the anti-Weyl fermions $\psi^m_\a$ from \BCpaper.}
from \massSweden.
The currents $\ce_P^m$, $\cf_P^{mn}$ and $\cX_P^\a$ are the Berends-Giele multiparticle polarizations of \BGBCJ,
\eqnn\multSYM
$$\eqalignno{
\ce_P^m &= {1\over k_P^2}\sum_{P=XY} \Big[\ce^m_Y(k_Y\cdot \ce_X) + \cf_X^{mn}\ce_Y^n + (\cX_X\g^m\cX_Y)
- (X\leftrightarrow Y)\Big]\,, &\multSYM\cr
\cf_P^{mn} &= k_P^m \ce_P^n - k_P^n \ce_P^m - \sum_{XY=P}\big(\ce_X^m \ce_Y^n - \ce_X^n\ce_Y^m\big)\,,\cr
\cX_P^\a & = {1\over k_P^2}\sum_{P=XY}k^n_{P}
\big[ \ce_X^m ( \g_n \g_m {\cal X}_Y)^\a - \ce_Y^m (\g_n\g_m \cX_X)^\a\big]\,,
}$$
starting with the single-letter gluon and gluino polarizations $\ce^m_i=e^m_i$, $\cX_i^\a=\chi_i^\a$
and field strength
$\cf_i^{mn} = k_i^m e^n_i - k_i^n e^m_i$.

The notation ${}{+}{\rm cyc}(P)$ instructs to add the cyclic permutations
of the letters in $P$ and $XY{=}P$ denotes the deconcatenations of
$P$ into non-empty words $X$ and $Y$. In addition, the momentum $k^m_P$ for a non-empty word $P=iQ$
is defined recursively by
$k_{iQ}^m = k^m_i + k^m_Q$ with $k^m_\emptyset = 0$, where {\it letters} are indicated by lower case and {\it words} by upper case.

\paragraph{Massless representation of massive superfields}
There are two distinct ways in which the massive superfields (labelled by $k$)
appearing
in the unintegrated pure spinor vertex operator can be represented
\refs{\massivevone,\UV,\Soares} in
terms of massless SYM superfields (labelled by $i$ and $j$): in the
{\it OPE} or {\it Berkovits-Chandia gauge} \UV. This rewriting is denoted by ${\underline
k}\to i,j$. More precisely,
\smallskip
\item{1.} ${\underline k}\to i,j$ in the OPE gauge:
\eqnn\OPEgauge
$$\eqalignno{
(\l B)_\a
&=- 2\ap \Bigl(ik_j^m(\g_m W_i)_\a V_j
+ ik_i^m (\l\g_m)_\a (W_i A_j)\cr
&\qquad\qquad{}-{1\over4}F_i^{mn}(\l\g^p\g^{mn})_\a A_j^p
- {1\over4}Q\bigl(F^1_{mn}(\g^{mn}A_j)_\a\bigr)\Bigr)&\OPEgauge\cr
(\l H^m)
& = -2i\ap\Bigl( k_j^n F_i^{mn}V_j
+ k_i^m (\l\g^n W_i)A_j^n + k_i^m Q(W_iA_j)\Bigr)\,,\cr
(C\l)^\a &= W_i^\a V_j\,,\cr
(\l F)_{mn}&=F_i^{mn}V_j\,,\qquad\qquad 2\ap k_i\cdot k_j = -1\,.
}$$
\item{2.} ${\underline k}\to i,j$ in the Berkovits-Chandia gauge:
\eqnn\massBC
$$\eqalignno{
(\l B)_{\a} &= (\g^{mnp}\l)_{\a}B_{mnp}\,,&\massBC\cr
(\l H_{m}) &= {3\over7}(\l\g^{np}D) B_{mnp}\,,\cr
(C\l)^\a &= {1\over 4}ik_q(\g^{qmnp}\l)^\a B_{mnp}\,,\cr
(\l F)_{mn} &= {1\over 16} \Big(7ik_{[m}(\l H_{n]}) + ik_q(\l \g_{q[m} H_{n]})\Big)\,,
}$$
with $2\ap k_i\cdot k_j = -1$ and
\eqn\BmnpBC{
B_{mnp} =
{1\over18}\ap^2\Bigl[(W_i\g_{ab mnp}W_j) k_i^a k_j^b
+ ik_j^{q}F^{i}_{q[m}F^{j}_{np]} +(i\leftrightarrow j) \Big]\,.
}\par
\noindent It was shown in \refs{\massivevone,\UV} that ${\underline k}\to k,
k{+}1$ implies a relation between massive and massless amplitudes given by
\eqn\affs{
A(1,P|{\underline k})\big|_{{\underline k}\to k,k{+1}} = -\langle C_{1|P,k,k{+}1}\rangle\,,\qquad
2\ap k_k\cdot k_{k{+}1}=-1\,,
}
where the superfields $C_{1|P,Q,R}$ are the scalar BRST invariants encoding the
$\ap^2$ terms of the massless string disk amplitudes, see \AFqToC.

\paragraph{Scalar BRST invariants}
The superfield expansions of the scalar BRST invariants
in terms of Berends-Giele currents follow from the recursion \partIcohomology
\eqnn\algoC
$$\eqalignno{
C_{i|j,k,l}&=M_i M_{j,k,l}\,,&\algoC\cr
C_{i|P,Q,R} &= M_i M_{P,Q,R} + M_i\otimes \big[C_{p_1|p_2 \ldots p_\len{P},Q,R} - C_{p_\len{P}|p_1 \ldots p_{\len{P}-1},Q,R} +
(P\leftrightarrow Q,R)\big]
}$$
where $M_i\otimes M_A := M_{iA}$, $M_P$ is the Berends-Giele
current associated to the unintegrated massless vertex operator and
\eqn\MABCdef{
M_{A,B,C} \equiv {1\over 3}(\l\g_m \cW_A)(\l\g_n \cW_B)\cF^{mn}_C +
{\rm cyc}(A,B,C)\,,
}
where $\cW^\a_P$ and $\cF^{mn}_P$ are Berends-Giele currents of the gluino and gluon
field strengths, for more details see the review \treereview.

The first few outputs of the recursion \algoC\ are given by
\eqnn\Cexs
$$\eqalignno{
C_{1|2,3,4} &= M_1 M_{2,3,4}\,, &\Cexs \cr
C_{1|23,4,5} &=M_1 M_{23,4,5} + M_{12} M_{3,4,5} - M_{13} M_{2,4,5}\,,  \cr
C_{1|234,5,6} &= M_1 M_{234,5,6} + M_{12}M_{34,5,6} + M_{123}M_{4,5,6} - M_{124}M_{3,5,6}\cr
&\quad{}- M_{14}M_{23,5,6} - M_{142}M_{3,5,6} + M_{143}M_{2,5,6}\,, \cr
}$$
and can be checked to be BRST closed using \partIcohomology
\eqnn\bvars
$$\eqalignno{
QM_P &= \sum_{P=XY}M_X M_Y\,,&\bvars\cr
Q M_{A,B,C} &=
\sum_{XY=A} \big[ M_{X}  M_{Y,B,C}
- (X\leftrightarrow Y)\big] + (A \leftrightarrow B,C) \,.
}$$
The relation between the scalar BRST invariants and the $\ap^2$ correction to the massless
disk amplitudes was discovered in \oneloopbb: writing the string disk amplitude as
\eqn\diskA{
A(P) = A^{\rm YM}(P) + \zeta_2 \ap^2 A^{F^4}(P) + {\cal O}(\ap^3)
}
it follows that $A^{F^4}$ can be expanded as
\eqn\AFqToC{
A^{F^4}(1,P) = \sum_{XYZ=P}\langle C_{1|X,Y,Z}\rangle,
}
while the precise permutations in the
inverse relation $\langle C_{1|P,Q,R}\rangle = \sum_S A^{F^4}(S)$ can be found
in the algorithm of \cdescent. Note that these BRST invariants also capture parts of genus-one
open-string amplitudes \refs{\MPS,\oneloopI}.

\newnewsec\Camps Massive field-theory amplitudes in superspace

\paragraph{Cohomology derivation of SYM amplitudes}
When all external states are massless, the field-theory SYM amplitudes could be determined using
the experimental observation that

{\narrower\smallskip\noindent \it two BRST-closed expressions with the same mass dimension and
featuring the same kinematic poles have
proportional component expansions.\smallskip}

\noindent Since BRST closed expressions are gauge invariant and supersymmetric under the application
of the pure spinor bracket \psf, and the SYM tree amplitudes can be obtained from the $\ap\to0$
limit of the tree-level open superstring amplitudes, SYM tree amplitudes
must be represented by a BRST closed expression in pure spinor superspace. Using the observation
above, any BRST closed expression of the same mass dimension and with the same kinematic poles must
yield the components of the SYM tree amplitudes.

This observation led to the idea that SYM tree amplitudes could be fixed by the cohomology of
pure spinor superspace \towards, which eventually came into fruition with
\nptMethod. We now conjecture that the same idea applies equally well to the
determination of {\it massive} field-theory amplitudes.

In this
programme, there is an implicit assumption used to propose a pure spinor superspace expression
reproducing the field-theory amplitude: as the starting point one uses superfields which are featured in the string
amplitude prescription of \psf. This reasoning led to the development of the multiparticle SYM
superfields inspired by OPEs \EOMbbs, and to the Berends-Giele interpretation \BGBCJ\ of the cohomology method of
\nptMethod. We expect similar developments for the massive superfields.

We are now going to showcase these ideas to determine the pure spinor superspace
expression for the massive field-theory amplitudes $A(P|{\underline n})$
involving one first-level massive and an arbitrary number of massless states.

\newsubsec\cohdersec From pure spinor superspace cohomology

We know that a single massive string state does not induce any kinematic poles,
therefore we will start with the pure spinor superspace expression for the massive field-theory
tree amplitudes
with a single massive state, denoted $A(1,P|{\underline n})$. Fortunately, the component
expansion of these amplitudes was determined in \massSweden\ as reviewed in
section~\revsec.

To find the pure spinor superspace expression that produces the Berends-Giele recursions of
\massSweden, the first step is to understand their kinematic pole structure.

The four-point amplitude $A(1,2,3|{\underline 4})$ has the poles $1/s_{12}$, $1/s_{23}$
and $1/s_{13}$ -- the same poles present in the scalar BRST invariant $C_{1|23,4,5}$ at multiplicity
five \oneloopbb. The pattern repeats at higher multiplicities: the poles of $A(1,P|{\underline n})$
are the same as the poles in $C_{1|P,n,n{+}1}$ due to their dependence on Berends-Giele currents.
Note that $A(1,P|{\underline k})$ and
$C_{1|P,k,k{+}1}$ have been recently related in a different context \refs{\massivevone,\UV}.

The second step in deriving a pure spinor superspace expression is the proposal of a BRST-closed
expression containing the same kinematic poles as outlined above.
The disk amplitude computed in \PSthreemass\ between two massless and one
first-level massive state
was simplified in \UV\ using BRST cohomology manipulations to
\eqn\tptaf{
A(1,2|{\underline 3}) = {i\over2\ap}\langle V_1 (\l\g_m W_2)(\l H^m_3)\rangle\,.
}
The three-point disk amplitude \tptaf\ turns out to be, under the definition in
section~1, proportional to  the massive field-theory
amplitude $A(1,2|{\underline 3})$.
Note that the expression in the right-hand side is BRST closed\foot{By abuse of
terminology, $\langle S\rangle$ is said to be
BRST closed when $QS=0$.}, as expected.
To see this, one uses the equations of motion \SYMBRST\ together with the pure spinor constraint
$(\l\g^m)_\a(\l\g_m)_\b = 0$.
Therefore, the simple pure spinor superspace expression \tptaf\
yields our starting BRST-closed expression and, by analogy
with the massless case reviewed above, we expect the higher multiplicity expressions to closely
follow the superspace structure of \tptaf. To generalize the three-point expression to a
BRST-closed expression of arbitrary
multiplicity containing the same kinematic poles as the scalar BRST invariants, it will be
convenient to define the following recursion:

\proclaim Definition. Pure spinor superfields $C^m_{1|P}$ for any non-empty word $P$ are given by
the following recursion
\eqnn\recCm
$$\eqalignno{
C^m_{i|j} &= M_i(\l\g^m \cW_j)\,,&\recCm\cr
C^m_{i|jk} &= M_i(\l\g^m \cW_{jk}) + M_i\otimes\big[C_{j|k}^m - C_{k|j}^m\big]\cr
C^m_{i|jPk} &= M_i(\l\g^m \cW_{jPk}) + M_i\otimes\big[C_{j|Pk}^m - C_{k|jP}^m\big]\cr
}$$
where $M_i\otimes M_P = M_{iP}$.

The first few cases of the recursion \recCm\ are given by
\eqnn\Cmexs
$$\eqalignno{
C_{1|2}^m &= M_{1}(\l\g^m \cW_{2})\,,&\Cmexs\cr
C_{1|23}^m &= M_{12}(\l\g^m \cW_{3})
+ M_{1}(\l\g^m \cW_{23})
- M_{13}(\l\g^m \cW_{2})\cr
C_{1|234}^m &=
 M_{1}(\l\g^{m}\cW_{234})
       	+ M_{12}(\l\g^{m}\cW_{34})
       	+ M_{123}(\l\g^{m}\cW_{4})
       	- M_{124}(\l\g^{m}\cW_{3})\cr&
       	- M_{14}(\l\g^{m}\cW_{23})
       	- M_{142}(\l\g^{m}\cW_{3})
        + M_{143}(\l\g^{m}\cW_{2})
}$$
with similar expressions at higher multiplicities. Notice the similarity with the corresponding
expansions of the scalar BRST invariants in \Cexs; in fact
$C_{1,P}^m$ can be obtained from those expansions by using the rule
$M_{Q,k,k{+}1}\to (\l\g^m \cW_Q)$. By analogy with the definition of the word
recursion in \Ree, one infers that $C^m_{1|P}$
is annihilated by proper shuffles
\eqn\shuCm{
C^m_{1|R\shuffle S}=0,\quad R,S\neq\emptyset,
}
where the shuffle product is recursively defined by $iA\shuffle jB = i(A\shuffle jB) +
j(B\shuffle iA)$ and $\emptyset\shuffle A=A\shuffle\emptyset =A$.
Moreover, it is easy to see that the recursion \recCm\ generates BRST closed
superfields containing two pure spinors.

Using the BRST closed expressions $C^m_{1|P}$ given above, the massive amplitude
$A(1,P|{\underline k})$ of arbitrary multiplicity
is proposed to be
\eqn\genAu{
A(1,P|{\underline k}) = {i\over2\ap}\langle C_{1|P}^m (\l H^m_k)\rangle\,.
}
Note that the shuffle symmetry \shuCm\ of $C^m_{1|P}$ implies, via the formula \genAu\ that
the massive amplitudes
$A(1,P|{\underline k})$ satisfy the Kleiss-Kuijf \KKref\ relations, in accordance with \massSweden.

By construction, the right-hand side of \genAu\ has the same kinematic poles as
the left-hand side. In addition, one can easily show that \genAu\ is BRST closed
using the equation of motion $Q(\l H^m_k)=(\l\g^m C_k\l)$ and the pure spinor constraint
$(\l\g_m)_\a(\l\g^m)_\b = 0$. Therefore we conclude that \genAu\ must yield component expansions
in terms of polarizations and momenta proportional
to the known components given by \APun. Indeed, using the $\t$ expansion
of $(\l H^m)$ from the appendix~\thetaexpsec\ and the identities
to extract component expansions automated in \PSS, we have explicitly verified \genAu\ up to $k=5$.

Therefore, the massive amplitudes \genAu\ have been derived from
the same pure spinor cohomology arguments as the SYM amplitudes of \refs{\towards,\nptMethod}.

\newsubsec\apdersec From massless $\ap^2$ amplitudes

The factorization of the massless $n{+}1$ amplitude on its first massive residue
was shown to be equivalent to the statement \UV
\eqn\Aufac{
A(1,P|{\underline k})\big|_{{\underline k}\to k,k{+}1} = -\langle C_{1|P,k,k{+}1}\rangle\,,
}
relating the massive amplitudes to the $\ap^2$ sector of the massless amplitudes,
in agreement with the earlier observation in \massivevone.
Equation \Aufac\ can be viewed as a consistency check due to unitarity, albeit written in
a slightly unconventional form. This statement was
explicitly verified \massivevone\ in terms of polarizations and momenta using the Berends-Giele construction
of $A(1,P|{\underline k})$ given in \APun\ on the left-hand side, and the component expansion
of the scalar BRST invariants available in \psswebsite.
In this case, the map ${\underline k}\to k,k{+}1$ is the component counterpart of the
superfield prescription \massBC, see \UV\ for the precise details.

If one has the $n$-point massive amplitude, then the map
${\underline k}\to k,k{+}1$ relates it to the kinematic expression governing the $\ap^2$
expansion of the massless amplitude at $n{+}1$ points. The more interesting direction would be to
derive the massive field-theory tree amplitudes starting from the massless string disk amplitudes;
that is, to invert the
factorization condition \Aufac. We will demonstrate below that the cohomology structure
of the pure spinor superspace allows us to do precisely that.

\paragraph{Inverting the factorization condition}
Since the pure spinor superspace expressions for both sides of the factorization
condition \Aufac\ as well as the superspace prescription of the map ${\underline k}\to k,k{+}1$
are known, we can exploit the simplicity of superspace to invert \Aufac: That is, we
want to arrive at the expression \genAu\ by
inverting the massless representation prescription ${\underline k}\to k,k{+}1$ given in
\OPEgauge\ and \massBC, starting from the right-hand side given in \algoC.

In order to do this, it will be convenient to rewrite the scalar BRST invariants in an asymmetric
manner.
One can show using a combination of equations of motion, pure spinor constraint and gamma matrix
identities that
\eqn\Calt{
C_{1|P,k,k{+}1} = C^m_{1|P}(\l\g^n \cW_k)\cF_{k{+}1}^{mn}
- Q\widehat M_{1|P,k,k{+}1}\,,
}
where the labels $k$ and $k{+}1$ are singled out to appear in different superfields.
In this equation, $\widehat M_{1|P,k,k{+}1}$ is given by the ghost-number two expression obtained
from the expansion of $C_{1|P,k,k{+}1}$ of \algoC\ and replacing
\eqn\replac{
M_{A,B,C}\to (\l\g^m \cW_A)(\cW_B\g_m \cW_C) + (\l\g^m \cW_B)(\cW_A\g_m \cW_C)\,.
}
For example,
\eqnn\Mhatex
$$\eqalignno{
\widehat M_{1|2,3,4} &= M_1(\l\g^m\cW_{2})(\cW_3\g_m\cW_4) + M_1(\l\g^m\cW_3)(\cW_{2}\g_m\cW_4)\,, &\Mhatex\cr
\widehat M_{1|23,4,5} &=
M_1(\l\g^m\cW_{23})(\cW_4\g_m\cW_5) + M_1(\l\g^m\cW_4)(\cW_{23}\g_m\cW_5) \cr
&+ M_{12}(\l\g^m\cW_{3})(\cW_4\g_m\cW_5) + M_{12}(\l\g^m\cW_4)(\cW_{3}\g_m\cW_5)\cr
&+ M_{31}(\l\g^m\cW_{2})(\cW_4\g_m\cW_5) + M_{31}(\l\g^m\cW_4)(\cW_{2}\g_m\cW_5)\,.
}$$
Since the pure spinor bracket annihilates BRST exact expressions we get
\eqn\CtoCm{
\langle C_{1|P,k,k{+}1}\rangle = \langle C^m_{1|P}(\l\g^n \cW_k)\cF_{k{+}1}^{mn}\rangle
=\langle C^m_{1|P}(\l\g^n \cW_{k{+}1})\cF_{k}^{mn}\rangle\,,
}
where the second equality follows from the BRST cohomology identity
\eqn\symC{
0 =\langle Q\big(C_{1,P}^m (\cW_k\g^m \cW_{k{+}1})\big)\rangle =
\langle C_{1,P}^m (\l\g^n \cW_k)\cF^{mn}_{k{+}1}\rangle
- \langle C_{1,P}^m (\l\g^n \cW_{k{+}1})\cF^{mn}_k\rangle\,.
}
The first equality represents the vanishing of BRST-exact expressions under the pure spinor
bracket \psf.
For the second equality, one uses $QC^m_{1|P}=0$, the equation of motion for $\cW^\a$ and the constraint $(\l\g_m)_\a(\l\g^m)_\b = 0$.

In the OPE gauge, the prescription ${\underline k}\to k,k{+}1$
for
the massless representation of the massive superfield $(\l H^m_k)$ is given by \OPEgauge
\eqnn\lHmsimpleA
$$\eqalignno{
(\l H^m_k) &= -2\ap\Bigl( ik_{k{+}1}^n F_k^{mn}V_{k{+}1}
 + ik_k^m (\l\g^n W_k)A_{k{+}1}^n + ik_k^m Q(W_kA_{k{+}1})\Bigr) &\lHmsimpleA\cr
&= -2i\ap\Bigl( (\l\g^m W_k)(k_k\cdot A_{k{+}1}) -
 F^{mn}_k (\l\g^n W_{k{+}1}) + Q(F_k^{mn}A_{k{+}1}^n) + k_k^m Q(W_kA_{k{+}1})\Bigr)\,,
}$$
where the second line follows from
$Q(F^{mn}_k A^n_{k{+}1}) = ik_{k{+}1}^n F^{mn}_k  V_{k{+}1} + ik_k^m(\l\g^n W_k)A^n_{k{+}1} - ik_k^n(\l\g^m W_k)A^n_{k{+}1}
+ F^{mn}_k (\l\g^n W_{k{+}1})$.

From \lHmsimpleA, we can formally rewrite the factorization ${\underline k}\to k,k{+}1$
in the reverse direction to obtain
\eqn\invertH{
2i\ap F^{mn}_k (\l\g^n W_{k{+}1}) = (\l H^m_k) + 2i\ap (\l\g^m W_k)(k_k\cdot A_{k{+}1})
+ Q\bigl((F_k^{mn}A_{k{+}1}^n) + k_k^m (W_kA_{k{+}1})\bigr)\,,
}
Finally, plugging \invertH\ into the BRST-equivalent expression \Calt\
of the scalar BRST invariant and using  and that $C_{1,P}^m$ is BRST closed
leads to
\eqnn\inverse
$$\eqalignno{
2i\ap \langle C_{1|P,k,k{+}1}\rangle &= 2i\ap \langle C_{1|P}^m F^{mn}_k (\l\g^n W_{k{+}1})\rangle &\inverse\cr
&= \langle C_{1,P}^m (\l H^m_k)\rangle + 2i\ap\langle C_{1|P}^m (\l\g^m W_k)(k_k\cdot
A_{k{+}1})\rangle + \langle C_{1|P}^m Q( \ldots)\rangle\cr
&= \langle C_{1|P}^m (\l H^m_k)\rangle \cr
&= -2i\ap A(1,P|{\underline k})\,,
}$$
where in the second line we used that $C_{1|P}^m(\l\g_m)_\a = 0$, integrated the BRST charge by
parts and used that $C^m_{1|P}$ is BRST closed
to obtain
$\langle C_{1|P}^m Q( \ldots)\rangle = -\langle \big(QC_{1|P}^m\big) ( \ldots)\rangle = 0$.

Therefore, inverting the massless representation map
${\underline k}\to k,k{+}1$ (which is equivalent to the factorization of the
massless string amplitudes on their first massive pole \UV)
yields the superspace expression of the massive
field-theory amplitude:
\eqn\fin{
\langle C_{1|P,k,k{+}1}\rangle \to - A(1,P|{\underline k})
}
This completes the (formal) derivation of the massive field-theory amplitude $A(1,P|{\underline k})$
from the kinematics $\langle C_{1|P,k,k{+}1}\rangle$ of the $\ap^2$ correction to massless
open-string disk amplitudes.

\newnewsec\concsec Conclusions

In this paper we derived, using BRST cohomology considerations,
a compact pure spinor superspace expression for the massive field-theory amplitudes $A(1,2,
\ldots|{\underline n})$. Furthermore, the same expression was also
derived from the $\ap^2$ correction to massless string amplitudes, as anticipated in \massivevone. The
successful application, in the massive case, of the central idea in \towards\ for massless
field-theory amplitudes leads us to conjecture that all massive field-theory amplitudes (as
defined in section~1) can be obtained by BRST cohomology considerations.

Furthermore, BRST cohomology manipulations in pure spinor superspace are powerful enough to lead one to
hope \massivevone\ that the expressions of
massive field-theory amplitudes with higher number of massive legs and/or higher mass levels can
be systematically obtained
from the known
massless disk amplitudes at higher $\ap$ orders. This paper gives evidence for the first step of this ladder, climbing
the rest of the way is left for future work.

\bigskip \noindent{\bf Acknowledgements:} CRM thanks Oliver Schlotterer for useful comments on the
draft, and Sitender Kashyap, Mritunjay Verma and Luis
Ypanaqu\'e for collaboration on related topics.

\appendix{A}{Theta expansion of massive superfields}
\applab\thetaexpsec

\noindent The $\t$ expansions of the massive superfields of the first massive level of the open
superstring have been determined
in \masstheta. In order to avoid problems due to different conventions -- especially due to
the convention $\p_m\to k_m$ used in the component expansions via \refs{\PSS,\FORM} --
we will rederive the expansions here in a streamlined manner.

\paragraph{Equations of motion and recursion}
With the definition
\eqn\Gmndef{
G^{mn} = -{1\over144}\bigl[(D\g^m H^n) + (D\g^n H^m)\bigr]\,,
}
one can show that the massive superfields satisfy \masstheta
\eqnn\eomtheta
$$\eqalignno{
D_\a G^{mn} &= -{1\over18}\p_p(\g^{pm}H^n)_\a - {1\over18}\p_p(\g^{pn}H^m)_\a\,, &\eomtheta\cr
D_\a B_{mnp} &= -{1\over18}(\g^{mn}H^p)_\a + {\ap\over18}\p_a\p_m\Big((\g^{an}H^p)_\a
-(\g^{ap}H^n)_\a\Big) + {\rm cyc}(mnp)\,,\cr
D_\a H^m_\b &=-{9\over2}G_{mn}\g^n_{\a\b}-{3\over2}\p_a B_{bcm}\g^{abc}_{\a\b}+{1\over4}\p_a
B_{bcd}\g^{mabcd}_{\a\b}\,.
}$$
Denoting by $[K]_n$ the component of the superfield $K$ of order $(\t)^n$, the Euler operator
$(\t D)$ satisfies $(\t D)[K]_n = n [K]_n$. Therefore multiplying \eomtheta\ by $\t^\a$ from the
left gives rise
to a recursion (note $\p_m \to k_m$)
\eqnn\eomthetarec
$$\eqalignno{
[G^{mn}]_k &= -{1\over 18k}\Big[k_p(\t\g^{pm}[H^n]_{k-1}) + k_p(\t\g^{pn}[H^m]_{k-1})\Big] &\eomthetarec\cr
[B_{mnp}]_k &= -{1\over18k}\Big[(\t\g^{mn}[H^p]_{k-1}) - \ap k_ak_m\Big((\t\g^{an}[H^p]_{k-1})
- (\t\g^{ap}[H^n]_{k-1})\Big) + {\rm cyc}(mnp)\Big]\cr
[H^m_\b]_{k} &={1\over k}\Big[-{9\over2}[G_{mn}]_{k-1}(\t\g^n)_{\b}
-{3\over2}k_a [B_{bcm}]_{k-1}(\t\g^{abc})_{\b}+{1\over4}k_a
[B_{bcd}]_{k-1}(\t\g^{mabcd})_{\b}\Big]\,.
}$$
starting with
\eqn\startrec{
[G_{mn}]_0 = g_{mn}\,\quad
[B_{mnp}]_0 = b_{mnp}\,,\quad
[H^m_\a]_0 = \psi^m_\a
}
of length dimensions $[g_{mn}]=0$, $[b_{mnp}]=1$ and $[\psi^m_\a]=\half$.
Using the recursion \eomthetarec\ yields the following $\t$ expansion for $\l^\a H^m_\a$ in the Berkovits-Chandia gauge:
\eqn\laHmtheta{
(\l H^m) = }
$${}(\l \psi^m)  -  {1 \over 4}\, (\l \g^{kmpqr}  \t) b_{pqr}
+  {3 \over 2}\, (\l \g^{kpq}  \t) b_{mpq}
-  {9 \over 2}\, (\l \g^{n}  \t) g_{mn} $$
$${} +  {1 \over 48}\Big[ (\l \g^{mnpqr}  \t) (\t \g^{np}  \psi^q) k_{i}^{r}
-  4 (\l \g^{npq}  \t) (\t \g^{mn}  \psi^p) k_{i}^{q}
-  2 (\l \g^{npq}  \t) (\t \g^{np}  \psi^m) k_{i}^{q} \Big]$$
$${} -  {1 \over 12}\Big [(\l \g^{npq}  \t) (\t \g^{nr}  \psi^p) k_{i}^{m} k_{i}^{q} k_{i}^{r} \ap 
 +  {3 \over 2}\, (\l \g^{n}  \t) (\t \g^{mp}  \psi^n) k_{i}^{p}
 +  {3 \over 2}\, (\l \g^{n}  \t) (\t \g^{np}  \psi^m) k_{i}^{p} \Big]$$
$${} -  {1 \over 576}\, (\l \g^{kmpqr}  \t) (\t \g^{p q r t u v k} \t) b_{tuv}
 -  {1 \over 32}\, (\l \g^{kmpqr}  \t) (\t \g^{p t k} \t) b_{qrt} $$
$${} +  {1 \over 96}\Big[ (\l \g^{kpq}  \t) (\t \g^{m p q s t u k} \t) b_{stu}
+  6 (\l \g^{kpq}  \t) (\t \g^{m s k} \t) b_{pqs}
-  12 (\l \g^{kpq}  \t) (\t \g^{p s k} \t) b_{mqs} \Big]$$
$${} -  {1 \over 48}\Big[ (\l \g^{k}  \t) (\t \g^{p q r} \t) b_{pqr} k^m
 -  {9 \over 2}\, (\l \g^{k}  \t) (\t \g^{q r k} \t) b_{mqr}
 -  {9 \over 2}\, (\l \g^{n}  \t) (\t \g^{q r k} \t) b_{nqr} k^m\Big] $$
$${} + {1\over48\ap}\Big[ (\l \g^{m}  \t) (\t \g^{n p q} \t) b_{npq}
 -  {9 \over2}\, (\l \g^{n}  \t) (\t \g^{m p q} \t) b_{npq}
 -  {9 \over2}\, (\l \g^{n}  \t) (\t \g^{n p q} \t) b_{mpq}\Big] $$
$${} -  {1 \over 32}\Big[ (\l \g^{kmpqr}  \t) (\t \g^{p q s} \t) g_{rs}
 - 4 (\l \g^{kpq}  \t) (\t \g^{m p r} \t) g_{qr}
 + 4 (\l \g^{kpq}  \t) (\t \g^{p s k} \t) g_{qs} k^m \ap  \Big]$$
$${} +  {1 \over 16}\Big[ (\l \g^{kpq}  \t) (\t \g^{p q r} \t) g_{mr}
 -  3 (\l \g^{n}  \t) (\t \g^{m q k} \t) g_{nq}
 -  3 (\l \g^{n}  \t) (\t \g^{n q k} \t) g_{mq}\Big] + {\cal O}(\t^4)$$
where a vector $k$ is written as an index if it is
contracted with a gamma matrix: for example $(\l\g^k\t)$ means $k_n (\l\g^n \t)$.
The $\t^4$ components are commented out in the \TeX\ file.
The $\t$ expansion of the other superfields are not needed in this paper but can
be easily generated from \eomthetarec.

\listrefs

\bye